\documentclass[epsfig,useAMS, usenatbib]{mn2e}

\usepackage{url}

\usepackage {graphics}
\usepackage{graphicx}
\usepackage {layout}

\newcommand{\kms}{{\rm km}\, {\rm s}^{-1}}
\newcommand{\Msun}{M_\odot}
\newcommand{\Vmax}{V_{\rm max}}
\newcommand{\beq}{\begin{equation}}
\newcommand{\eeq}{\end{equation}}
\newcommand {\hi} {H\,{\small I}\,}

\newcommand\aj{AJ}
\newcommand\apj{ApJ}
\newcommand\apjl{ApJ}
\newcommand\apjs{ApJS}
\newcommand\mnras{MNRAS}
\newcommand\araa{ARA\&A}

\title[Mass-dependent formation of disk galaxies]{On the morphologies, gas fractions, and star formation rates of small galaxies}

\author[Kaufmann et al.]
{Tobias Kaufmann$^{1}$ \thanks{E-mail: tobias.kaufmann@uci.edu}, Coral Wheeler$^{1}$ and James S. Bullock$^{1}$ 
\\$^1$ Center for Cosmology, Department of Physics and Astronomy, University of California, Irvine, CA 92697}

\begin{document}

\pagerange{\pageref{firstpage}--\pageref{lastpage}} \pubyear{} 

\maketitle

\begin{abstract}

We use a series of  N-body/smoothed particle hydrodynamics simulations
and  analytic arguments  to  show  that the presence   of an effective
temperature  floor in  the interstellar medium   at $T_F \sim 10^4$  K
naturally explains    the tendency  for  low-mass   galaxies to  be  more
spheroidal, more gas rich, and  less  efficient in converting  baryons
into   stars than  larger galaxies.   The    trend arises because  gas
pressure   support becomes  important   compared  to angular  momentum
support in small dark  matter haloes.  We  suggest that  dwarf galaxies
with  rotational velocities $\sim 40$ km  s$^{-1}$ do not originate as
thin discs, but rather are born as  thick, puffy systems.  If accreted
on to larger   haloes,   tenuous dwarfs of   this    kind will  be  more
susceptible  to gas   loss  or tidal transformation   than scaled-down
versions of  larger  spirals.    For a  constant  temperature   floor,
pressure   support becomes less   important  in large  haloes, and this
produces  a tendency for  massive  isolated galaxies  to have  thinner
discs  and  more  efficient star formation   than  their less  massive
counterparts, as observed.

\end{abstract}


\section{Introduction}

It is well established that small galaxies have longer star formation
time scales than their larger cousins (e.g.  Hunter \& Gallagher 1985;
van  Zee 2001; Kauffmann et  al.   2003; Brinchmann et al. 2004; Geha et   al.  2006).   Dwarf
galaxies in the  field have  higher  specific gas  fractions (van  Zee
2001; Geha et  al.  2006) and  are morphologically thicker (Dalcanton et al. 2004; Yoachim \&
Dalcanton 2006)  than  luminous late-type  galaxies.  It is  common to
discuss supernova feedback or radiative feedback as a means to explain
these trends (e.g.  Dekel \& Silk 1986; White \& Frenk 1991; Kauffmann
et al. 1993; Bullock et al.  2000; Mayer et al. 2001a; Somerville 2002;
Dekel \& Woo 2003;  Kravtsov et al. 2004; Read et
al. 2006;  Stinson  et al.  2006),  yet   the origin  of  the observed
relations between galaxy properties  and their total mass remains open
for debate.

In  this paper we investigate the  simple systematic effect that a gas
temperature floor, $T_F$, has on  the star formation efficiencies, gas
fractions,  and morphologies of galaxies as  a function of dark matter
halo  mass.  We show that even  a moderate effective temperature floor
of $T_F \simeq 10^4$ K, as might arise naturally  in the presence of a
photo-ionizing  background,   produces  many of  the  general   trends
observed.  Specifically,    pressure   support  becomes    dynamically
comparable to angular momentum support in small dark matter haloes, and
this  causes  dwarf galaxies   to be systematically   puffier and less
efficient in converting gas into stars than their larger counterparts.

In  what follows  we 
explore a range of  values $T_F = (1.5  - 5) \times 10^4$ K, in
order to  qualitatively mimic  real physical  effects that may  act to
pressurise  gas  within   galaxies.   These may  include   inefficient
cooling,  heating   by  an  internal or external
ultraviolet   (UV) background,   supernova
feedback,  turbulent pressure,  and cosmic-ray heating,  among others.
Of course, within a fully realistic galactic interstellar medium (ISM)
we  expect molecular cooling to  produce cold clouds embedded within a
warm, pressurised medium (e.g. Mckee  \& Ostriker  1977; Yepes et  al.
1997;  Springel  \&  Hernquist  2003;  Robertson et  al.  2004, 2006)
Internal and external heating sources   and cooling within the  medium
should produce a quasi-stable  system  with an effective   temperature
that  acts to stabilise the galaxy.   Unfortunately, the nature of the
coupling between various   energy sources and  the   background ISM is
poorly understood.  Rather than  attempt to model these processes, we
 use $T_F$ as a phenomenological proxy.

Our approach is perhaps most directly relevant to  the situation of an
 ionizing background field  that acts to prevent cooling below
$\sim  10^4$ K in  warm galactic  gas.  It  is  well  known that a UV
background  can suppress galaxy formation    altogether in very  small
galaxies $\la 30 \, \kms$  (Efstathiou 1992; Thoul \& Weinberg  1996;
Quinn et  al. 1996; Gnedin  2000;  Hoeft et al.  2006; Crain et al. 2007)  and that  this
effect can have important implications for the baryonic mass fractions
and  overall abundance of the smallest  galaxies in the universe (e.g.
Bullock et  al.  2000;  Benson et al.   2002; Strigari et al. 2007).  In a
recent examination,   Hoeft et
al. (2006)  used cosmological simulations with cooling and star formation
 to show that gas  at typical
galactic  densities will have  equilibrium temperatures $\sim (1-3)
\times 10^4$ K, and that this temperature is fairly insensitive to
the normalisation of the UV flux.  This heating strongly suppresses the
baryon  fraction in   haloes smaller   than  $\sim  20 \, \kms$  in  their
simulations.  Our   investigations  focus on  galaxies   that are just
above this  scale.  Specifically, we  explore the morphology and
star  formation efficiency of galaxies  that  are large enough to accrete
warm gas 
but small enough to be dynamically affected by warm gas pressure.

We note that the effect of a finite  temperature floor on dwarf galaxy
formation was  discussed in  a semi-analytic  context by  Kravtsov  et
al. (2004), who   used  the idea   to motivate  models for  low   star
formation rates in small galaxies; and by Tassis et al. (2006), who used
a similar  model to investigate the  gas fraction and mass-metallicity
relationships in dwarf  galaxies.     Taylor \& Webster (2005) 
studied star formation within equilibrium dwarf galaxies by modeling
H$_2$ cooling within a thermally-balanced $\sim 10^4$ K medium and
derived lower limits on self-regulated star formation rates in dwarfs in
this context.

In the  next section we present an analytic investigation into
the importance of  baryonic pressure support  compared to angular momentum
support as a function of virial mass and gas temperature and show that
pressure  support should become  dynamically important in dwarf-galaxy haloes.  In  \S 3 we 
use the  N-body/smoothed particle hydrodynamics 
(SPH) code \textsc{Gasoline} (Wadsley et al. 2004) to explore galaxy
formation with a  range  of temperature floors and  halo   masses.  We
present results on   galaxy  disc thickness, gas  fractions,  and star
formation rates as a function of galaxy circular velocity.  We reserve
\S 4 for discussion and \S 5 for conclusions.

\section{Analytical Expectations}

The  standard  analytic  approach  for  calculating  galaxy  sizes and
morphologies within dark matter haloes assumes that the  gas cools to a
temperature  well below the  halo virial temperature,  $T_g \ll T_{\rm
v}$.  As a  result, the thermal pressure support in the gas is
small compared to its angular momentum support.
A {\em thin}
disc  of star-forming   material is the    natural  outcome (Fall   \&
Efstathiou 1980,   Blumenthal  et  al.     1986).   This  thin-disc
configuration is  taken to be the starting  point for galaxy formation
(and star formation) in most models, including those that
model  dwarf galaxies  (e.g.  Kauffmann et  al.   1993; Somerville  \&
Primack 1999; Benson et al. 2002; Somerville 2002; Mastropietro et al.
2005; Mayer et  al.  2006; Gnedin et al.   2006; Dutton et al.  2007).
Here we stress that the thin disc approximation should break down 
in small haloes where $T_{\rm v} \ga T_g \sim 10^4$ K.  
In these cases, the pressure
support radius becomes comparable to the angular momentum support
radius and we expect a thick  morphology.
What follows is
a simple analytic investigation aimed at
quantifying the halo mass scale of relevance.  A more rigorous numerical 
approach is given in the next section.

Consider dark matter haloes of mass $M_{\rm v}$ with virial radii
defined\footnote{In our definition, the virial radius
contains an  average mass density equal  to
$360$ times the matter  density of the universe and we adopt
 $\Omega_m = 1  -
\Omega_{\Lambda} = 0.27$ and with $h = 0.7$.} such that
$R_{\rm v} \simeq 113  \, {\rm kpc} \, (M_{\rm v}/10^{11} \Msun)^{1/3}$.
We assume that halo density profiles 
are well approximated by the NFW fit (Navarro, Frenk, \& White 1996):
\begin{equation}
\rho(x) = \frac{\rho_s}{x (x + 1)^2} \, ,
\end{equation}
where $x \equiv r/r_s$ and the scale radius, $r_s$, is determined from
the halo concentration   parameter  $c_{\rm v} \equiv    R_{\rm  v}/r_s$.
We adopt the relation $c_{\rm v} = 10 \,
(M_{\rm v}/10^{11} \,  \Msun)^{-0.086}$, which is appropriate
for a $\sigma_8 = 0.75$ $\Lambda$CDM cosmology 
(Bullock et al. 2001a; Macci{\`o} et al. 2007). 
Given $M_{\rm v}$ and $c_{\rm v}$, 
the circular velocity curve, 
$V_c^2(r) = G M(r)/r$, is determined by the integrated mass profile.
For our adopted relation,
the circular velocity peaks at a maximum value
$V_{\rm max} \simeq 71 \kms (M_{\rm v}/10^{11} \, \Msun)^{0.3}$.

In the  standard   thin-disc  scenario,  the gas obtains specific
angular momentum that is similar to that of the dark matter,
which is often  characterised   by a  dimensionless spin
parameter (Peebles 1969) defined as  $\lambda \equiv j \, |E|^{1/2} \,
G^{-1} \,  M_{\rm v}^{-3/2}$, where $G$  is Newton's  constant and $j$
and $E$ are the specific angular momentum and energy of the halo, respectively.
Simulated CDM haloes typically  have $\lambda \sim 0.03$
with a 90  \% spread between $0.01  - 0.1$ (e.g.  Barnes \& Efstathiou
1987;  Bullock  et  al.  2001b;   Macci{\`o} et   al.  2007).   

It  is
straightforward to show that if the gas  cools and contracts without
angular momentum loss to form a
thin, angular  momentum  supported exponential   disc, the disc  scale
radius is given by (Mo, Mao, \&  White 1998; hereafter MMW)
\beq
R_d = \lambda \, R_{\rm halo} \, f(c,\lambda, m_d). 
\label{eqn:mmw}
\eeq 
Here we assume that the gas falls in from a radius $R_{\rm halo}$,
defined to be either the virial radius, $R_{\rm v}$, or the
``cooling radius'', $R_{\rm c}$, depending on which one is smaller
$R_{\rm halo} = {\rm min}(R_{\rm c},R_{\rm v})$.  By introducing
the cooling
radius (White \& Frenk 1991) we account for the expectation that
hot gas in the outskirts of massive haloes will not have had time to 
cool since the halo formed.  For simplicity, we adopt\footnote{
 This approximation 
was given by Maller \& Bullock (2004) for haloes with
$V_{120} \ga 1$.  Below this scale, the virial radius sets $R_{\rm halo}$.}
$R_{\rm c} = \, 129 \; {\rm kpc} \, V_{120}^{-1/4}$, where
$V_{120}$ is the halo maximum circular velocity in units of
120 km  s$^{-1}$.  
The function $f$ ($ \sim 1$) in Equation \ref{eqn:mmw} 
contains information on the halo profile shape and
contraction from baryonic infall, and depends on
the initial halo concentration
$c \equiv R_{\rm halo}/R_{\rm s}$
and the disc mass, $m_d$, in units
of the total mass within $R_{\rm halo}$.

\begin{figure}

\includegraphics[scale=0.4]{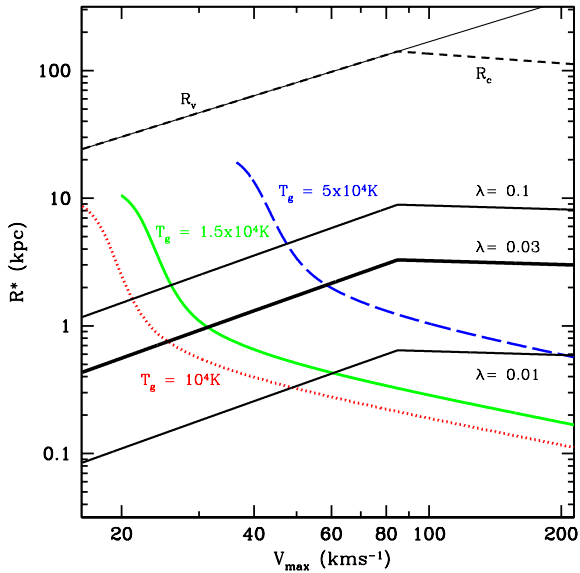}
\caption{Galaxy radius $R^*$ versus halo $V_{\rm max}$ for various models
for galaxy assembly. $R^*$ is defined as the radius that contains $26\%$ of the
galaxy mass and is equivalent to the  disc scale radius for a thin exponential disc.
 The solid lines that increase with decreasing halo size show
$R^*$ for a pressure-supported isothermal gas with temperature floors as labelled.
The straight lines that break at $\Vmax \sim 90 \, \kms$ show $R^*$ for an angular momentum supported thin disc
for three different halo spin parameters as labelled.  The break occurs when
the infall radius transitions from the virial radius (R$_{\rm v}$)
to the cooling radius ($R_{\rm cool}$).
Note that the pressure support radius becomes comparable to (and even larger than) the
angular momentum support radius in small galaxies. \label{fig1}}
\end{figure}

The three solid lines labelled with $\lambda$ values in Figure
\ref{fig1} show $R_d$ calculated in the standard thin-disc framework
as a function of the (initial) halo $\Vmax$ value.  From top to bottom
the lines assume spin parameters $\lambda = 0.1$, $0.03$, and $0.01$.
We have used the fitting formula from MMW  $f(c, \lambda,
m_d)$ with $m_d = 0.1$.  We plot the scale radius as a function of the
initial, uncontracted halo $\Vmax$ in order to facilitate the
following comparison.

\begin{table*}
\begin{center}
\caption{Simulated Galaxies \label{table-ic}}
\begin{tabular}{||l|cccc|lll||}
\hline  \hline 
\, \, \, \, \, (1) & (2) & (3) & (4) & (5) & \, (6) & \, (7)& \, (8) \\
\, \, \, Name   & $\Vmax$    & \,  $c_{\rm v}$ \,   & $\, M_{\rm v}$   
& $T_{F}$   & \, $R_{\rm d}$  & \, $z_{\rm d}$ &  \, V$_{2.2}$  \\
&	\, [$\kms$] \,  & \, \, \,   & \, [$10^{10}$ M$_\odot$] \,  &  \, [$10^4$ K] \,  & [kpc] &  [kpc] & [$\kms$] \\
\hline 
D24 		&24 	&   	14 & $0.29 $ & 1.5  & 0.37	& 0.08 & \, 37 \\
D24\_g 		&24 	&   	14 & $0.29 $ & 1.5  & 0.35	& 0.19 & \, 35 \\
D28 		&28 	&	13 & $0.47 $ & 1.5  & 0.55	& 0.09 & \, 43 \\
D28\_g 		&28 	&	13 & $0.47 $ & 1.5  & 0.56	& 0.18 & \, 41 \\
D41 		&41 	&	12 & $1.6  $ & 1.5  & 0.92	& 0.10 & \, 65 \\
D41\_g  	&41 	&	12 & $1.6  $ & 1.5  & 1.02	& 0.16 & \, 65 \\
D41\_W 		&41 	&	12 & $1.6  $ & 3.0  & 0.61	& 0.13 & \, 64 \\
D53 		& 53 	&	11 & $3.8  $ & 1.5  & 1.32	& 0.12 & \, 84 \\
D53\_g	 	& 53 	&	11 & $3.8  $ & 1.5  & 1.34	& 0.15 & \, 84 \\
D53\_W 		&53 	&	11 & $3.8  $ & 3.0  & 1.05	& 0.16 & \, 82 \\
D53\_H 		& 53 	&	11 & $3.8  $ & 5.0  & 0.77	& 0.18 & \, 84 \\
G74 		& 74 	&	10 & $11.0 $ & 1.5  & 1.6	& 0.14$^{\dagger}$ & \, 115 \\
G74\_g	 	& 74 	&	10 & $11.0 $ & 1.5  & 2.2$^{*}$	& 0.03$^{*}$ & \,  92 \\
G148 		& 148 	&	8  & $100.$ & 1.5  & 1.88	& 0.22$^{\dagger}$ & \, 209 \\
G148\_g		& 148 	&	8  & $100.$ & 1.5  & 1.9$^{*}$	& 0.02$^{*}$ & \, 217 \\
G168\_gHRLS	& 168 	&	11  & $115.$ & 3.0 & 1.94	& 0.13 & \, 213 \\

\hline 
\end{tabular}
\end{center}
{\small (1) 
Names labelled  ``\_g'' refer to pure gas runs without star formation whereas names with ``\_W'' and ``\_H'' signify ``warm'' or ``hot'' temperature floors and
``HRLS'' refers to the Milky Way model described in Kaufmann et al. (2007). 
(2-4) The listed values for $\Vmax$, $c_{\rm v}$ and $M_{\rm v}$ are {\em initial} halo parameters.  
(5) $T_F$ is the imposed temperature floor.  (6-8) The last three
columns list parameters measured in the final galaxy, where $R_d$ and $z_d$ are the exponential scale length and
scale height, respectively, and
$V_{2.2}$ is the  rotational velocity in the gas measured at $2.2$ $R_d$.  Disc scale heights marked with ($\dagger$) are
artificially large as a result of numerical heating.  Gas disc parameters marked with ($*$) are derived from discs which have become Toomre unstable; thin, but spatially irregular.}
\end{table*}

Consider now the galaxy radius that would result from pressure support 
in an idealised, spherically
symmetric system consisting of a gravitationally sub-dominant
gas.  Assume that this gas has no angular momentum but reaches a
temperature of $T_g = T_{\rm F}$ within an extended NFW halo
of virial temperature $T_{\rm v}$. We emphasize that we
 will calculate the pressure support radius for ``cold'' gas at $T_g = T_{\rm F}$, the lowest temperature the gas can reach within our assumptions.
For this approximate calculation, we neglect baryonic contraction.
We will define the virial temperature
in analogy with an isothermal gas
of temperature $T$ and its associated speed of sound
$c_s$ (e.g. Maller \& Bullock 2004):
\beq
T = 10^{4} K \left(\frac{c_s}{11.5 \, \kms}\right)^2.
\label{eqn:temp}
\eeq
Specifically, 
$T_{\rm v}$ is set by using $\Vmax/\sqrt{2}$ for $c_s$
in Equation \ref{eqn:temp}.

The equilibrium gas profile, $\rho_g(r)$, 
will be set by a competition
between the isothermal gas pressure, $P_g = c_g^2 \rho_g$, and the 
gravitational potential.
If we assume that the gravitational force is dominated by the
NFW halo potential, the 
hydrostatic force balance equation
\begin{eqnarray}
\frac{c_g^2}{\rho_{g}} \frac{d\rho_{g}}{dr} & = & \frac{-V^2(r)}{r} \, , 
\end{eqnarray}
can be rewritten in terms of the dimensionless radial parameter $x=r/r_s$ as
\begin{eqnarray}
 \frac{1}{\rho_g} \frac{d\rho_{g}}{d x} & = & - \eta \, h(x) \, .
\label{eqn:Hydro}
\end{eqnarray}
Here,  $\eta \equiv T_{\rm v}/T_{\rm F}$ 
parameterises the relative strength of the
halo gravity and the thermal pressure of the gas and
$h(x) = 9.26 x^{-2} \left[\ln(1+x) - x/(1+x) \right]$.  Note that in the
limit of large $\eta$ ($T_{\rm v} \gg T_{\rm F}$), 
the gas profile will be centrally concentrated with 
a large negative derivative, and a
 negligible pressure-support radius.  
More generally, solving Equation \ref{eqn:Hydro} for $\rho_g$ yields
\begin{equation}
\rho_{g}(x)=\rho_0 \exp \left[-9.26 \, \eta \, \Big{(}1 - \frac{ \ln[1+x]}{x} \Big{)} \right],
\label{eqn:profile}
\end{equation} 
where the normalisation parameter $\rho_0$ sets the gas density at $x=0$.
It is clear that in small
haloes with $\eta \sim 1$, the gas profile can extend to $x \sim 1$ or
$r \sim R_{\rm v}/c \sim 0.1 R_{\rm v}$, which is comparable in size to
the angular momentum support radius (Eq. \ref{eqn:mmw}).

The    thick-solid and dotted   lines in   Figure \ref{fig1}  show  a  more explicit
comparison   for various gas floor 
temperatures:  $T_{\rm F} = 1$,   $1.5$, and $5
\times 10^4$ K.  We have   used Equation \ref{eqn:profile} to  compute
the  radius $R^*$ that  encloses $26\%$  of  the pressure-supported
galaxy mass as a function of halo  $\Vmax$.  This radius is
analogous to  the scale  radius for an  exponential disc, which
contains   $26  \%$ of the  rotationally   supported disc mass.  While
relatively  unimportant in large  Milky-Way-size  haloes ($\Vmax \sim  200 \, 
\kms$),   we   see that   pressure   support  should 
dominate in shaping galaxy morphologies and gas distributions in small haloes.
The expectation is that small galaxies will be intrinsically 
puffier than large galaxies,  even in the   absence of environmental influences.   

The  scale where pressure becomes important  compared to rotation will
naturally  depend on the   temperature  floor of  the  gas and  on the
intrinsic spin.  For $T_{\rm F} = 1.5 \times 10^4$ K and $\lambda =0.03$, we
expect  the effect to become very  important for dwarf-size haloes with
$\Vmax \la 35 \, \kms$.  If the temperature floor  is high, $T_{\rm F} = 5
\times 10^4$  K, then the effect could  be important even in $\sim 100
\,  \kms$  haloes if they have inhabit the low-spin tail of the
distribution, $\lambda = 0.01$.  If galaxies form with  the range of spins
expected ($\lambda  \sim   0.01-0.1$)  then  we  would predict  a   range  of
morphologies (from puffy to thin discs) {\em at fixed $\Vmax$} as long
as the temperature floor is roughly the same from galaxy to galaxy.
 Though we do not explore galaxy formation in very small halos ($ \la 20 \kms$) in the rest of this paper, it is interesting to note that we expect
the morphologies of the smallest 
objects to be essentially spheroidal, with initially
extended gas profiles.  The stellar sizes of these objects will likely be
much smaller than the gas extent, as high-densities will be required for
star formation. This might provide an explanation for why the smallest
galaxies (dwarf spheroidals) are always dispersion supported systems.

The simple, spherical, model we have just explored was primarily
 designed to guide expectations.  
Stronger results are presented in the next section, where we use  3D  hydro-dynamical
simulations to investigate the  effect  of  a reasonable   gas
temperature  floor on morphologies, gas  fractions, and star formation
rates in small galaxy haloes.

\section{SPH Simulations}

We use the parallel TreeSPH code \textsc{Gasoline} (Wadsley et
al. 2004), which is an extension of the pure N-Body gravity code
\textsc{Pkdgrav} developed by Stadel (2001). It includes artificial 
viscosity using the   shear reduced  version   (Balsara 1995)  of  the
standard Monaghan   (1992)  implementation.  \textsc{Gasoline}  uses a
spline   kernel  with   compact  support   for  the softening  of  the
gravitational and SPH quantities. The  energy equation is solved using
the asymmetric   formulation,  which is shown  to  yield  very similar
results compared  to the entropy  conserving formulation but conserves
energy  better (Wadsley  et  al.  2004).  The code  includes radiative
cooling  for     a  primordial   mixture   of   helium   and  (atomic)
hydrogen.  Because  of the lack  of  molecular cooling and metals, the
efficiency of our cooling functions  drops rapidly below $10^4$ K.
The   lack of molecular cooling    is unimportant in our investigation
because  we enforce temperature  floors $T_{\rm F} \ge 1.5  \times 10^4$
 K.

We investigate runs with and without star formation.
The adopted star formation
recipe  is similar to that described   in Katz (1992).  Specifically, a
gas particle may  spawn  star particles if  i)  it is in  an over-dense
region; ii) it is  cool,  with $T = T_{\rm   F}$;  and iii) it has   a
density greater than a critical threshold,  $\rho_g > \rho_{\rm SF}$.
In practice, the critical star formation  density is the most important
parameter.  In our primary simulations   we use $\rho_{\rm SF} =  2.5
\times 10^{6}  \Msun \, {\rm    kpc}^{-3}$, but explore a case with
$\rho_{\rm SF}$ increased by a factor of 100 in \S 4.  
Once a gas particle is eligible for spawning stars, it
does  so based on   a  probability distribution  function with  a star
formation efficiency factor $c^*$ that is tuned to match the Kennicutt
(1998) Schmidt Law for the Milky Way and M33-size  disc described in
Kaufmann et al. (2007). The mass of  gas particles decreases gradually
as they spawn more star particles.  After its mass has decreased below
$10\%$ of its  initial value the gas particle  is removed and its mass
is re-allocated  among the neighboring gas particles.   Up to six star
particles are then created for each gas particle in the disc.  We note
that Stinson et al.  (2006) have implemented  a similar star formation
recipe, although  they include an  allowance  for supernova feedback
effects using a subgrid, multi-phase model based on blast waves.

\subsection{Initial conditions} 

We simulate   15 isolated systems  with masses   spanning the scale   of  dwarf
galaxies   to large spirals,  with    {\em initial}  maximum  circular
velocities that range  from $\Vmax = 24$ to  $148 \, \kms$.  We also
present results from an older 
simulation of a  Milky-Way size galaxy ($\Vmax = 168 \, \kms$) 
that was originally discussed in Kaufmann et al. (2007).  
Haloes  are
initialised as  spherical equilibrium NFW  profiles using  the methods
outlined  in   \cite{Stelios}   and  we  use  the   mass-concentration
relationship discussed  in \S 2 to set  the profile parameters.   Table 1
lists the specific parameters  used in each  simulation and provides a
reference name for each run.

We initialise a fraction of the total halo mass, $f_{\rm b}=0.1$, as a
hot  baryonic component with the  same radial distribution as the dark
matter and impose a temperature profile such that the gas is initially
in hydrostatic  equilibrium with an adiabatic  equation of state.  For
all of our fiducial  models we choose $\lambda_g =  0.03$ for  our gas
spin parameter,  defined in analogy  with the  halo spin as $\lambda_g
\equiv j_{g} |E|^{1/2} G^{-1} M_{\rm v}^{-3/2}$.  Here, $j_{g}$ is the
average specific angular momentum of the gas, $E$  and $M_{\rm v}$ are
the total energy and mass of the {\em halo}.  

The specific angular momentum  distribution of the  gas is  assumed to
scale linearly with the cylindrical distance from the angular-momentum
axis of  the halo, $j  \propto r  ^{1.0}$.  This  choice is consistent
with values found for  dark  matter haloes within  cosmological  N-body
simulations (Bullock et al.  2001b). For simplicity, we initialise the
dark matter particles with no net angular momentum.

The hot gaseous halo  is sampled with $N_g  = 10^5$ particles  and the
dark matter halo  with $N_{\rm dm}  =  2 \times  10^5$ particles.  The
force resolution is set  to be a fixed fraction  of the simulated halo
virial radius, $f_{\rm   res}   = 0.002 R_{\rm v}$.      These choices
correspond to cases where  numerical losses of angular momentum become
small  (Kaufmann  et  al.    2007).   A  detailed description  of  our
initialisation method and results on  the evolution of 
a Milky Way-size galaxy is presented in Kaufmann  et
al. (2007).

As seen in Table   1, our new   simulations sample six  halo  circular
velocity scales  spanning those of dwarfs  to large spirals.  For each
$\Vmax$,  we ran a case with  and  without star  formation and  used a 
conservatively low temperature floor $T_F = 1.5 \times 10^{4}$ K. 
We performed two additional $\Vmax =  53 \, \kms$ simulations with
a  ``warm'' and  ``hot'' temperature floors at  
$T_F  = 3$ and   $5 \times 10^4$ K,
respectively, and simulated a second $\Vmax =  41 \, \kms$ case with
$T_F = 3 \times 10^4$ K.  Higher temperature  floors were not explored
in  the  smaller haloes   because  galaxy  formation is suppressed  all
together if  $T_F \sim  T_{\rm v}$.\footnote{Of  course, in  a more
complicated  scenario, feedback from star formation  may be the source
of the a high effective gas temperature,  in which case a dwarf galaxy
could form, and  subsequently lose its gas once  it is heated to  $T_g
\sim T_{\rm v}$  (see Stinson et al. 2007).}

The galaxies were evolved for  $5$ Gyr, but  we  find that the  global
results stabilise after $3$ Gyr (see Figure 6 below).  The two largest
``gas-only'' runs without star formation (G74\_g and G148\_g) produced
gas   discs  that   were  unstable  to  their  own   self-gravity  and
became clumpy.  This is not  too surprising given the low $T_F$
adopted,  and is consistent with  previous claims that substantial heating of the interstellar medium is needed   to stabilise  disc  galaxies  (e.g.  Robertson  et
al. 2004).   While we  have   listed radially and  vertically  average
``disc'' properties  for these unstable cases  in  Table 1,  we do not
include these systems in relevant figures  below.  In order to present
results for a large, stable, pure gas disc, we use the $\Vmax = 168 \,
\kms$  simulation from  Kaufmann et  al. (2007)  with  $T_F = 3 \times
10^4$ K.  This system had  a slightly larger  spin parameter than  the
rest of our runs ($\lambda_g = 0.038$).

We note that  our  final
galaxies have  larger maximum   circular  velocity scales than   their
initial haloes because of the effects of 
baryonic contraction (compare the
second and last columns in  Table 1).   
For example,  our   D41  series produces galaxies  that  are
comparable in rotation speed  to dwarf irregular galaxies like the
Large Magellanic Clouds at $\sim 60 \, \kms$.  Our smallest (D24) 
runs produce galaxies that are large enough 
($\sim
35 \, \kms$) to be included in the Geha et al. (2006) sample
of SDSS dwarfs.  Our larger galaxies,  G74 and
G148, produce systems that are comparable to M33 
($\sim 100 \, \kms$) and the Milky Way ($\sim 200 \, \kms$).

Finally, we address  our initial conditions
in  light of
the idea of ``cold flows'' put forward by Kere{\v s} et al. (2005) and
Birnboim  \&  Dekel (2003).   These   authors  find that  gas   is not
shock-heated to the virial temperature in small haloes $M_{\rm v} \la 10^{11}
\Msun$, but rather is accreted   as ``cold'' material, with $T_g  \la
10^{5}$ K.   Our models are  not  strongly at  odds with this picture.
Specifically, the gas within our small haloes cools very quickly to the
temperature  floor  and indeed falls into  the  central region  in its
``cold'' phase.

\begin{figure}
\includegraphics[scale=0.475]{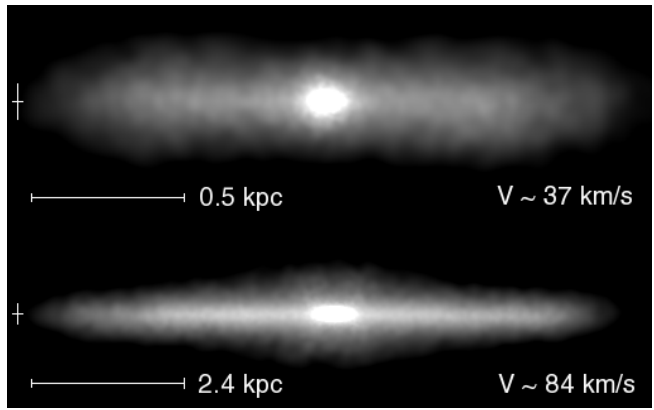}
\caption{Edge-on views of a simulated dwarf galaxy (D24, upper) and 
a more massive galaxy (D53, lower).  The gray-scale maps the 
projected stellar density.  The disc of the larger galaxy is clearly thinner than the
disc of the dwarf. The vertical bars indicate twice the softening length used in the simulations. \label{fig2}}
\end{figure}

\begin{figure}
\includegraphics[scale=0.4]{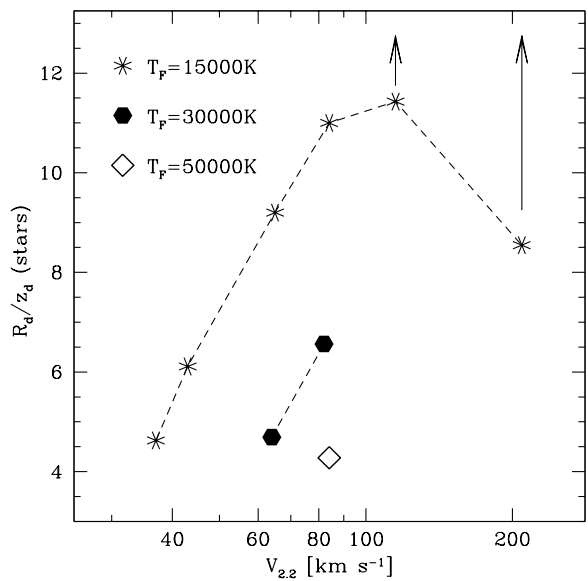}
\caption{Galaxy disc ``thinness'' as a function of circular velocity for our star formation runs.  
Here $V_{2.2}$ is the rotational velocity  of the gas measured at $2.2
R_{\rm   d}$ for  each  disc.  Symbol  types  correspond  to different
temperature floors as indicated.   Discs are thicker in smaller  haloes
and    for larger temperature floors.   The   arrows indicate that the
stars in the largest   galaxies  have  been artificially    thickened  by numerical heating.
\label{fig3}}
\end{figure}

\begin{figure}
\includegraphics[scale=0.4]{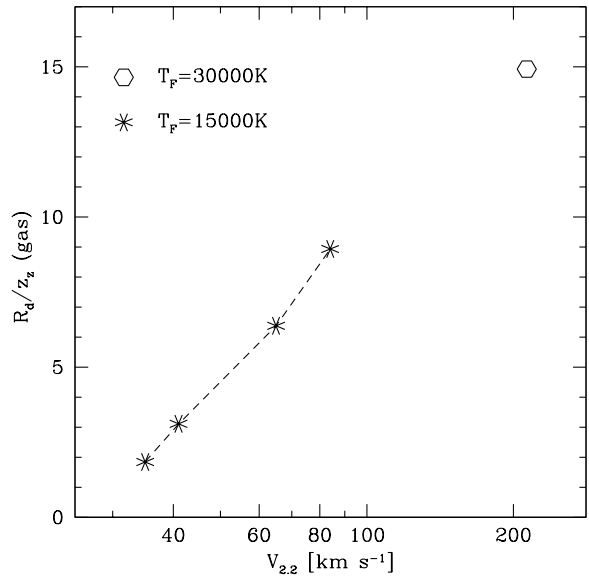}
\caption{The disc ``thinness''  as a function of galaxy rotational velocity
at $2.2 \, R_d$ for our  pure gas runs.  
Symbol types correspond to
different temperature floors as indicated.
The morphological trend is
quite similar to that obtained for our runs with star formation (Figure \ref{fig3}). \label{fig4}}
\end{figure}

\subsection{Results 1: morphological trends}

The upper  and lower panels of  Figure \ref{fig2} illustrate the final
projected stellar density for the D24 and  D53 runs.  It
is evident that  even the  mild temperature  floor,  $T_{\rm F} =  1.5
\times 10^4$ K has resulted in a very thick disc for the small,
$\sim 35 \, \kms$ galaxy,
while the larger system, $\sim 85 \, \kms$, is closer to a standard
thin disc.  

We note that both of the systems in Figure \ref{fig2} have formed
substantial central bulge components.  These central stellar mass concentrations
are likely an artifact of our simple initial conditions, which assume
a centrally-concentrated  NFW profile for the hot gas.  We return to this
potential shortcoming in \S 4.

In order to provide a more quantitative comparison between runs, we have estimated a
disc scale length, $R_{\rm d}$, and scale height $z_{\rm d}$, for each
galaxy.  We have explored  several methods for quantifying $R_{\rm d}$
and $z_{\rm d}$ and  find that our  overall results change very little
between methods.   Because of the large  bulge component, our galaxies
are   not well  described by   a   single exponential  surface density
profile.   Instead  of adopting  a   degenerate two-component fit,  we
define $R_{\rm  d}$ by simply measuring  the radius where the face-on stellar
(or gas)  surface density  drops to  $0.03$  (or  $e^{-3.45}$) of  its
central value, and define this radius to be equal to $3.45 R_{\rm d}$.
This method is generally insensitive to the details of the functional
form of the final disc and probes a radial
range that is well sampled with star (or gas) particles.
The vertical scale height, $z_{\rm d}$
is determined by fitting an exponential profile to the projected, edge-on
surface density profile at a projected radius equal
to $R_{\rm d}$ (this avoids the bulge region).  The fit is typically good
out to vertical scales as large as $z \sim 3 \, z_d$ above the disc. 
The measured values for each simulation
are listed in Table 1.  

Figure \ref{fig3} presents 
the galaxy ``thinness'' ratio ($R_{\rm d}/z_{\rm d}$) as a function of 
the  (edge-on) {\em gas}  rotational velocity, $V_{2.2}$, measured
at $2.2 R_{\rm d}$.  The points indicated by stars correspond to the
$T_{\rm F} = 1.5 \times 10^4$ K cases, while the solid hexagons and open diamond
show $T_{\rm F} = 3$ and $5 \times 10^4$  K, respectively.  It is clear
that at a fixed $T_{\rm F}$ small galaxies are naturally thicker than
large galaxies.  This simple prescription reproduces well the general
trend found by Yoachim \& Dalcanton (2006) for edge-on galaxy thickness
as a function of circular velocity (e.g. their Figure 5).

We  note that the   two largest galaxies in
Figure \ref{fig3} are  artificially   thickened (as indicated by   the
arrows).  Both  force softening and two-body  heating are important in
these systems.  The force softening acts  as an artificial pressure at
small scales (Bate \& Burkert 1997)  and this effect becomes important
in the two most massive galaxies  G74 and (especially) G148, where the
vertical extent of the  disc becomes comparable to softening parameter
($f_{\rm res} \simeq 0.25$ and $0.5$ kpc, respectively).  Furthermore,
because the   massive galaxies  are  more efficient  in spawning  star
particles, the ratio of the average star  particle mass to dark matter
particle mass is higher in the larger galaxies (the ratio is one order
of  magnitude higher in G148   compared to D24).  We therefore  expect
that  our    measured $z_d$ values   are  {\em  over-estimates} of the
vertical scale heights compared to what would  have been achieved with
higher resolution   simulations  (which  would   be  numerically  very
expensive).    These numerical problems act preferentially to thicken the
larger galaxies.  Therefore, the general trend with decreasing thickness
see between $\sim 40$ and $100 \, \kms$ should be robust.

\begin{figure}
\includegraphics[scale=0.4]{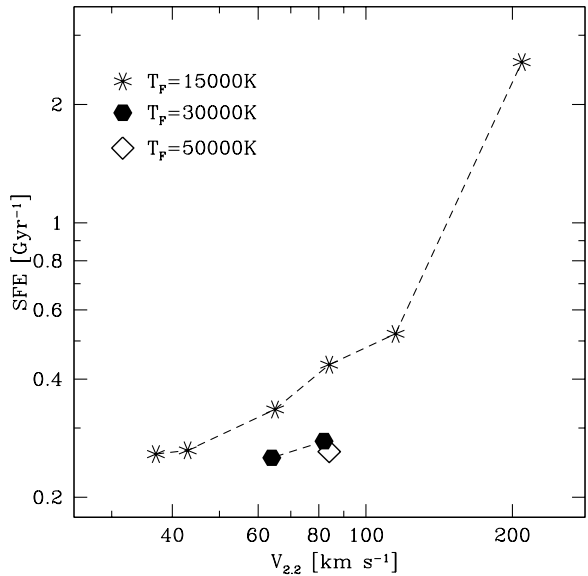}
\caption{Star formation efficiency (see text) as a function of
the galaxy rotational velocity. 
Symbol types correspond to
different temperature floors as indicated. \label{fig5}}
\end{figure}

Figure \ref{fig4}  shows that the same morphological  trend is seen in our
our pure gas runs without star formation.  We do not plot the massive
G74\_g and G143\_g galaxies.  As mentioned above, the gas became so cold in
these runs ($T_F = 1.5 \times 10^4$ K) 
that the  final  discs fragmented into  thin,
clumpy, irregular systems.  The  open hexagon at  $V_{2.2} = 213 \, \kms$
is simulation G168\_gHRLS, where HRLS refers to the Milky Way model described in Kaufmann et al. (2007).  This system is hot enough to be stable, with $T_F = 3
\times   10^{4}$ K, and   ends up  as  very  thin  disc.  Overall, the
agreement between  our pure  gas runs and   those with star  formation
suggests    that the correlation   between  disc thinness and circular
velocity should hold, and is largely independent of uncertainties 
associated with star formation.

\subsection{Results 2: gas fractions and star formation}

In the previous  section we showed that  galaxies formed within  small
haloes tend to be thicker than those formed within large haloes.  Figure
\ref{fig5}  shows that the   star  formation efficiency (SFE) in   our
simulated galaxies also varies as a function  of $V_{2.2}$.  We define
SFE $= \dot{m}_{*}/m_{\rm  g}$, where $\dot{m}_*$   is the star
formation rate  and $m_{\rm g}$  is the gas  associated with the
galaxy.  Specifically, $m_{\rm g}$ is  defined to be the mass of gas
that is both cold ($T = T_F$)
and no  longer infalling.  It is  evident from Figure  \ref{fig5} that
our dwarf  galaxies are less efficient  in turning gas into stars than
are  larger galaxies,   as expected.   Moreover,   at  fixed  circular
velocity, the efficiency is reduced for higher ISM temperature floors.

\begin{figure}
\includegraphics[scale=0.4]{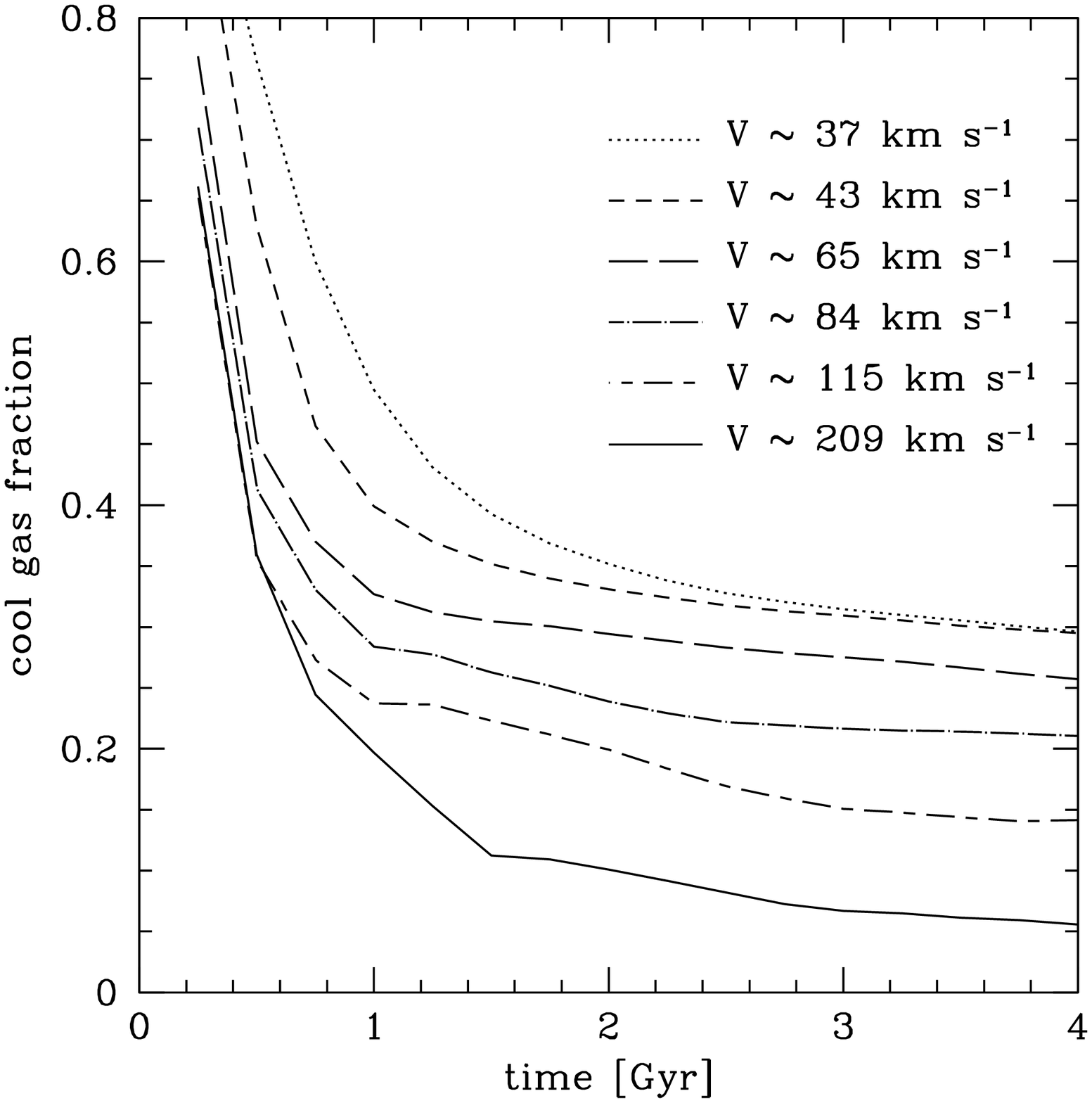}
\includegraphics[scale=0.4]{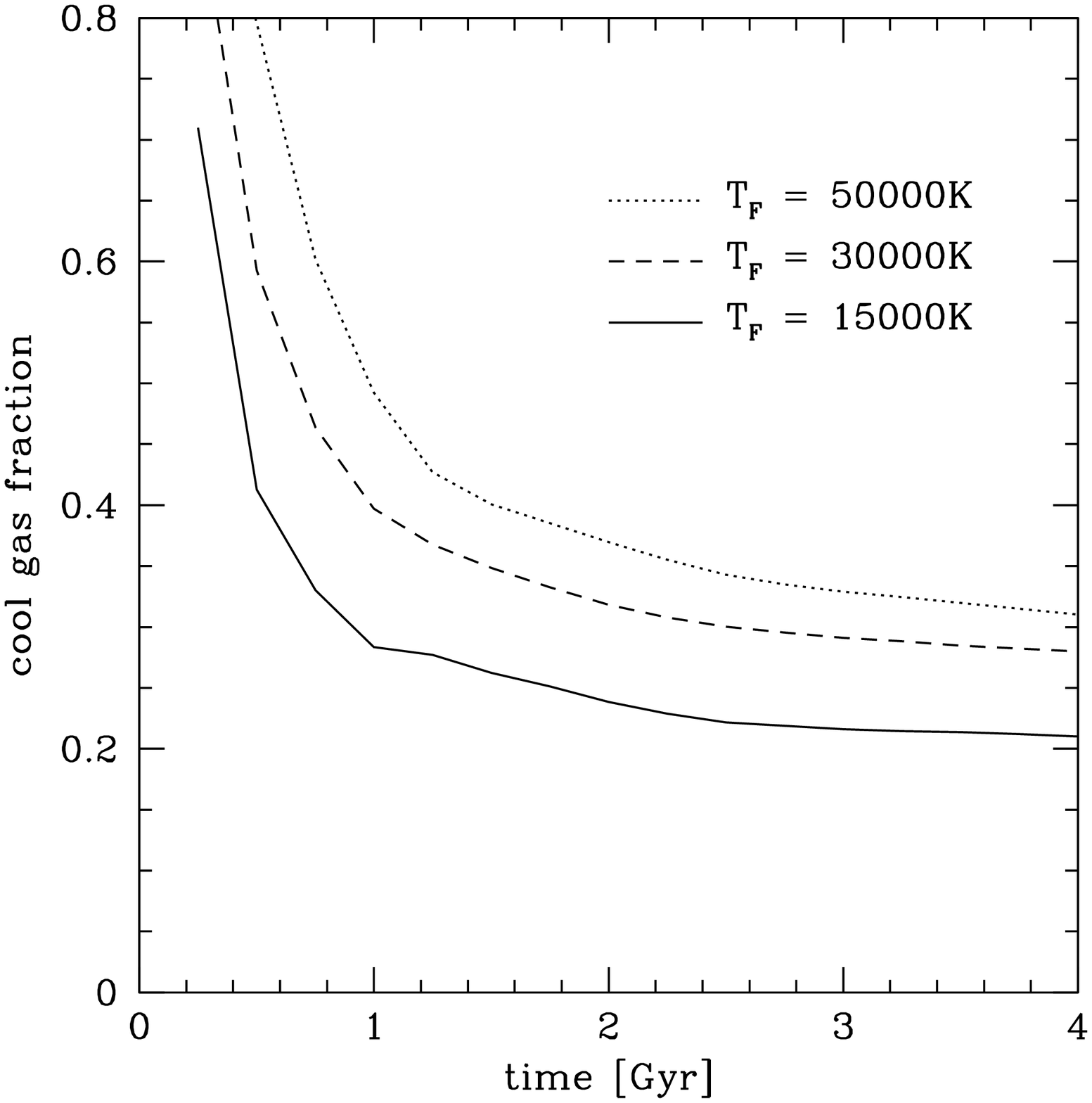}
\caption{The fraction of cool gas in galaxies as a function of time.
The {\em upper  panel} shows runs with different halo masses ($V_{2.2}$ listed) and a fixed
ISM temperature floor $T_F = 1.5 \times 10^4$ K.
The {\em lower panel} shows a series of models with the same halo mass
(the D53 series) but different $T_F$ values. \label{fgas_VT} } 
\end{figure}

Another observationally-oriented measure of the efficiency of
star formation 
is the cool gas fraction,  $m_{\rm g}/(m_{\rm
g}+m_{*})$.
The upper and lower panels of Figure \ref{fgas_VT} show the evolution 
of the cool gas with time in our galaxies.  We find that the cool gas
fraction approaches a constant after $\sim 3$ Gyr.  This corresponds to
the time when the infall of new cool gas reaches an equilibrium with the
rate that gas is being converted into stars in the galaxy.
The upper panel shows the evolution for galaxies with different (final)
circular velocities ($V_{2.2}$) at a 
fixed ISM temperature $T_F = 1.5 \times 10^4$ K.  The
lower panel
shows our D53 series for three values of $T_F$.  
Larger systems end up with lower gas fractions, as do systems with
decreasing ISM temperatures.
Figure \ref{fig7} shows the same data sliced at a fixed time (3 Gyr) plotted as a function
of $V_{2.2}$.  Clearly, the gas fractions are higher in smaller galaxies, as 
expected.

\section{Discussion}

\begin{figure}
\includegraphics[scale=0.4]{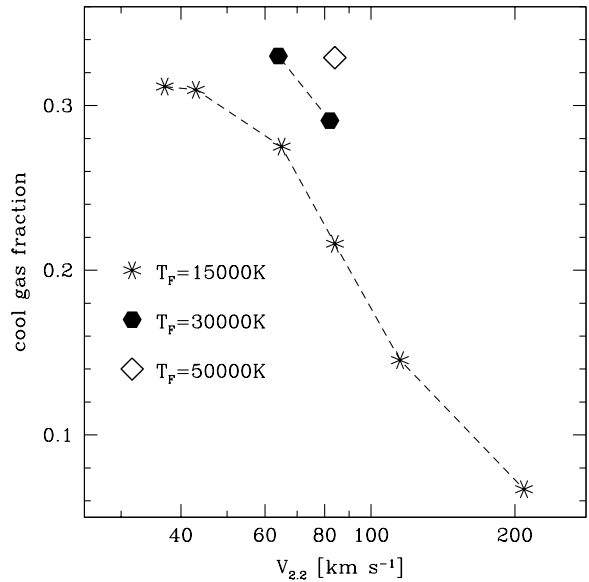}
\caption{The fraction of cool gas in galaxies as a function of circular velocity.
Symbol   types   correspond   to  different   temperature   floors  as
indicated. \label{fig7} }
\end{figure}

Geha et al. (2006) looked at a  sample of 101 extremely low luminosity
dwarf galaxies selected from the Sloan Digital Sky Survey, and found a
trend for dwarfs  to be systematically much more  gas rich than giants
(e.g.  Geha et  a.  2006, Figure  3).   The  results presented  in the
previous section (Figure \ref{fig7})  show encouraging agreement  with
the  observed  trend.  However,  Geha et  al.    find an   average gas
fraction  in dwarfs of  $\sim 0.6$, and our  simulated dwarfs have gas
fractions that are lower $\sim 0.3$.    The comparison may be
even worse than it appears because the observations constrain only the
neutral \hi  fraction.  At $T  \sim  10^4$ K, however, the  difference
between neutral and total fraction is expected to  be small.  While we
regard the predicted {\em  trend}  between gas fraction and   velocity
scale  as  the most   robust   aspect  of  this   work,  it  is  worth
investigating whether simple  adjustments might bring our results into
closer agreement  with the   data.\footnote{Note, however, that  these
observations   possibly  underestimate   the stellar  masses   because
contributions from  extended,    low    surface  brightness    stellar
populations  are    likely  to    be  missed   (M.   Blanton,  private
communication; or see Roberts \& Haynes 1994; van Zee 2001).}

\begin{figure}
\includegraphics[scale=0.4]{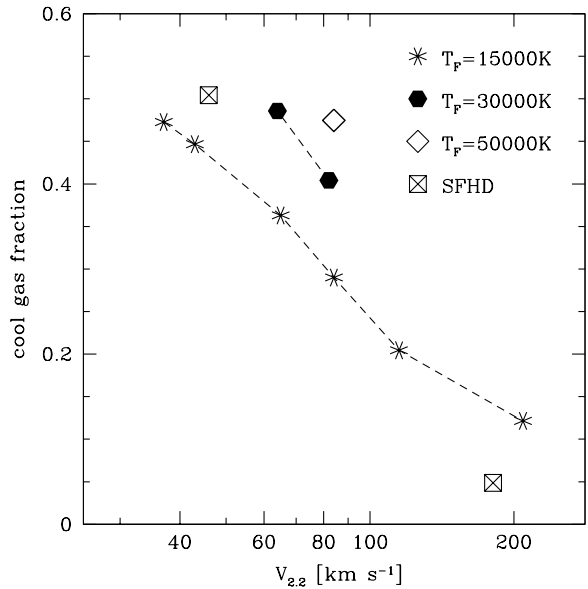}
\caption{
Stars, hexagons, and open diamond show
the cool gas fraction in the {\em disc} for the fiducial series of simulated galaxies
(bulge stars have been removed from this
measurement, see text for details).  
As before, the symbol types indicate the temperature
floor assumed.  The {\em crossed squares} show the galaxy gas fraction
in runs with an increased star formation 
density threshold $\rho_{\rm SF} \rightarrow 100 \rho_{\rm SF}$. For those runs the bulge stars have been included in the measurements.
\label{fig8} }
\end{figure}

An obvious problem with our simulations  is that our galaxies all form
pronounced bulges (see Figure 2).   As discussed elsewhere, the bulges
may be artifacts of  our  simplistic initial conditions, which  assume
pure NFW  profiles  for the  hot gas   haloes   (Kaufmann et al.,   in
preparation; see also  Hansen  \& Sommer-Larsen 2006; Mastropietro  et
al. 2005). In Figure \ref{fig8}, we have attempted to account for this
obvious shortcoming  by plotting the  {\em   disc} gas  fraction as  a
function of circular  velocity for the  same set of galaxies shown  in
Figure  \ref{fig7}.   Here,   we  have  removed the   central  stellar
component in computing the disc gas  fraction, $m_{\rm g}/(m_{\rm g} +
m_{*  \,  {\rm  disc}})$.   Specifically, we   do not  include central
stellar material with a  spherically-averaged density larger than $2.3
\times 10^{9}$  M$_{\odot}$   kpc$^{-3}$,  and  this   eliminates  the
bulge-like nuclear structures in our galaxies.  As  seen by the stars,
hexagons, and  open diamond in  Figure 8, these ``disc'' gas fractions
are more in line with observations.

Of  course,  another clear uncertainty  in any galaxy formation 
simulation is  star formation.  The most
important parameter in our  prescription is the density
threshold  for  star formation, $\rho_{\rm  SF}$.  Observationally, we
can constrain only the relationship  between the projected gas density
and star formation rate (Kennicutt 1998), however, because the density
threshold in  our prescription is  three-dimensional, we are left with
the freedom to explore its parameter space.\footnote{We refer the reader to Kravtsov (2003) for an interesting
theoretical discussion on the origin of the Schmidt-Kennicutt
relation.}  The squares with 
crosses in Figure  \ref{fig8} show  the result of  two  runs (D24\_d and
G148\_d) with  $\rho_{\rm SF}$ set at  $100$  times our fiducial value
(to $2.5 \times 10^{8}$ M$_{\odot}$ kpc$^{-3}$). Note that in order to
more directly compare with the fiducial runs, we have not
excluded bulge stars in  these  points.
As might be expected, the increased threshold produced a much higher gas
fraction for  the dwarf galaxy  compared to  the standard case (Figure
7).  It  also  produced a  smaller disc than  the fiducial  run, but a
similar axis ratio.  Unlike the small galaxy, the  gas fraction in the
G148\_d system is quite similar to that in the fiducial run (Figure 7)
because the gas was able to  become quite dense  and form stars.  Like
the  fiducial  Milky Way-size  galaxy, this  system  also  sits on the
Kennicutt relation.  We note, however, that  unlike the fiducial case,
G148\_d produced $\sim 10$ very dense star clusters in the final disc.
Numerical scattering off  of these clusters  dramatically affected the
star particle orbits, and artificially increased  the scale height of
the final disc.

Elmegreen  \& Parravano (1994), Blitz \&   Rosolowski (2004, 2006) and
Wong \& Blitz (2002) pointed out the relation between the gas pressure
in the midplane of the galaxy and the ratio of H$_{2}$ versus \hi, and
related also the star  formation efficiency. The  gas pressure in  the
midplane $P/k_B$ in our simulations has  values ranging from $4 \times
10^4$ to $8 \times 10^5$ cm$^{-3}$ K in agreement with the findings of
Blitz \& Rosolowski (2006), but does  not show a strong evolution with
galaxy mass.  In the simulations  the pressure stabilises the gas well
above  the  midplane for  the dwarfs,  but  not for  the  more massive
galaxies, making  the midplane pressure less  meaningful for the dwarf
galaxies.

While it is not the aim of
this paper to reproduce the observations in detail, we conclude that
there are several physically plausible effects that can eventually 
lead to a more complete understanding of the gas fractions in small
systems.  Overall, the
agreement between the predicted and observed {\em trends}
is quite encouraging.

\section{Summary and Conclusions}

We have used SPH simulations and an analytic
discussion to argue that many of the observed changes in
galaxy properties as a function of their rotation speed arise 
naturally because of the increased importance of ISM pressure in
small haloes. Our main results may be summarised as follows:

\begin{itemize}

\item An effective ISM temperature floor at $\sim 10^4$ K produces a pressure
support radius that is comparable to the angular momentum support radius in
dwarf galaxies, $\sim 40 \, \kms$.  This suggests that most small galaxies are
not formed as thin discs, but rather are born as thick, puffy systems.

\item For a constant temperature floor, pressure support becomes less important in 
large haloes, and this naturally produces a tendency for 
massive, isolated galaxies
to have thinner discs than their less massive counterparts, as observed.

\item The morphological trend produces related trends in star formation efficiency: dwarf galaxies are predicted to have longer star formation time scales than larger galaxies.  Similarly, galaxy gas fractions decrease with circular velocity, as observed.

\end{itemize}

The expected morphological trend seems to be  fairly independent of star
formation details (compare Figures \ref{fig3} and \ref{fig4}).
While relations of this kind are difficult to quantify observationally
for a large number of galaxies, 
Yoachim  \& Dalcanton
(2006) used a sample of $34$  late-type,  edge-on,
undisturbed disc galaxies to 
show that more  massive  galaxies are generally thinner than  less massive
galaxies (see their Figure 5).  They find radial to vertical axis-ratios for
dwarf galaxies as low as $\sim 3$, in agreement with results presented in
our Figure \ref{fig3}.

More clues  to the nature of galaxy  formation on small  scales can be
gained from the  population  of  dwarf spheroidal galaxies    (dSphs).
Unlike equally-faint   dwarf irregular  galaxies (dI), dSphs   are gas
poor.    The fact that gas-poor dwarfs   are exclusively  found in the
proximity of a  luminous neighbor (Geha et  al.  2006) encourages  the
notion that tidal forces  and ram pressure  stripping act to transform
dI type galaxies into  dSphs (Gunn \& Gott  1972;  Lin \& Faber  1983;
Moore \& Davis  1994; Mastropietro et al.  2005;  Mayer  et al. 2001a,
2001b, 2002,  2006, 2007).   Typically, models  aimed at  testing this
transformation hypothesis initialise a  {\em thin disc} within a small
dark matter halo and investigate how tides  or ram-pressure affect the
galaxy as it  falls into a  larger host.  

Mayer et al.  (2006, 2007) used  Smoothed Particle Hydrodynamics (SPH)
simulations   to show  that  the combined   effects  of tides and  ram
pressure can convert discy dwarfs to  gas-poor spheroidal systems, but
only   if a heating source  is  imposed to keep  the  gas in the dwarf
extended and  hot  at a   temperature  of  $\sim 2.5   \times   10^4$K
(L.  Mayer, private  communication).   As we  have  shown, temperature
floors of this magnitude inhibit the formation of thin discy dwarfs in
the field.  If a  puffy dwarf  of the  kind we  expect falls into  the
potential well  of a larger galaxy with  an extended  hot gas halo, it
should  be  quite  susceptible    to   gas loss  and     morphological
transformation.  Trends of this kind  do seem to  be broadly in accord
with some observations (e.g. Lisker et al. 2007, Geha et al. 2006).
In another investigation,  Mastropietro et
al.  (2005)  used  N-body simulations to  study the  transformation of
discy  dwarfs to spheroidal dwarfs.  They  had  some success, but were
unable  to reproduce the observed  fraction of  spheroidal dwarfs with
negligible rotational support.   Our results suggest that field dwarfs
are likely to be born thicker than  typically assumed and are therefore more
susceptible to kinematic transformations.

It has long been recognised that galaxy formation must become increasingly
inefficient in dark matter haloes from the scale  of big spirals
to small dwarfs (White
\& Reese 1978; Klypin et al. 1999; Moore  et al. 1999; Strigari et al.
2007).    Our work suggests that it
is  difficult to avoid  the suppression of galaxy formation efficiency
in small haloes.  Shallow potential  wells naturally give rise to puffy
galaxies with  long star formation  time scales.  This  situation makes
them    more  susceptible  to  other    feedback  effects and external
influences that may act to suppress star  formation even further.  
Small galaxies  are also systematically  more metal poor  than
larger systems (e.g. Tremonti et al. 2004). The
lower  efficiency of  star  formation  may also  explain the  observed
mass-metallicity relation  without the need for   strong winds (Tassis et al. 2006).  
For
example, Brooks et al.  (2007) recovered in their cosmological simulations
 the observed relation only by including an allowance for ISM
heating (see also Ricotti \& Gnedin 2005). 

A natural extension of this work would  be a more detailed modeling of
the coupling between various energy sources and the ISM. Ideally, this
would include a star formation prescription based on molecular cooling
and radiative transfer for the treatment of the feedback from stars to
the ISM  -  allowing to overcome  the  simple  proxy of a  temperature
floor.

\section*{Acknowledgments}

It is a pleasure to thank James Wadsley, Joachim Stadel and Tom Quinn for making
\textsc{Gasoline} available to us. The numerical simulations were performed on the
zBox2      supercomputer   at      the     University   of    Z\"urich
(\url{http://www.zbox2.org})  and the IA64  Linux  cluster at the  San
Diego Supercomputer Center.  We thank Doug  Potter, Joachim Stadel and
Ben Moore for building the  zBox2 and allowing  us to run parts of the
simulations  on it.  We would  like to  thank Stelios Kazantzidis  for
providing   a   code to  generate  isolated  dark   matter  haloes.  We
acknowledge useful and  stimulating discussions with  Michael Blanton,
Julianne Dalcanton, Marla Geha,  Andrey Kravtsov, Ariyeh Maller, Lucio
Mayer, Brant Robertson, Greg Stinson, Liese van Zee, Betsy Barton and Louis Strigari.   We  also thank the anonymous referee for valuable comments. This work was
supported by the Center for Cosmology at UC Irvine.



\end{document}